\documentstyle[12pt]{article}
\renewcommand{\theequation}{\thesection.\arabic{equation}}

\def\a{\begin{eqnarray}}
\def\b{\end{eqnarray}}
\def\0{\nonumber}
\def\ba{\begin{array}}
\def\ea{\end{array}}

\def\ep{{\epsilon}}
\def\al{{\alpha}}
\def\lm{{\lambda}}

\def\tl{{\widetilde L}}
\def\tm{{\widetilde M}}

\def\hs{\hat \sigma}
\def\res{{\rm res}}

\def\th{{\theta}}
\newlength{\extraspace}
\newlength{\extraspaces}
\newcounter{dummy}
\newcommand{\ai}{
\addtocounter{equation}{1}
\setcounter{dummy}{\value{equation}}
\setcounter{equation}{0}
\renewcommand{\theequation}{\thesection.\arabic{dummy}\alph{equation}}
\begin{eqnarray}
\addtolength{\abovedisplayskip}{\extraspaces}
\addtolength{\belowdisplayskip}{\extraspaces}
\addtolength{\abovedisplayshortskip}{\extraspace}
\addtolength{\belowdisplayshortskip}{\extraspace}}
\newcommand{\bj}{
\end{eqnarray}
\setcounter{equation}{\value{dummy}}
\renewcommand{\theequation}{\thesection.\arabic{equation}}}
\def\d{{\partial}}

\def\tt{\widetilde t}

\newcommand{\ddt}[1]{{\partial \over \partial t_{#1}}}

\newcommand{\ddg}[1]{{\partial \over \partial g_{#1}}}

\newcommand{\bac}{\begin{array}{c}}
\newcommand{\bacc}{\begin{array}{cc}}
\newcommand{\baccc}{\begin{array}{ccc}}
\newcommand{\barcl}{\begin{array}{rcl}}
\newcommand{\bacccc}{\begin{array}{cccc}}
\newcommand{\baccccc}{\begin{array}{ccccc}}
\newcommand{\baccccccc}{\begin{array}{ccccccc}}
\newcommand{\barclcrcl}{\begin{array}{rclcrcl}}
\newcommand{\bacl}{\begin{array}{cl}}
\newcommand{\bal}{\begin{array}{l}}
\newcommand{\bacll}{\begin{array}{cll}}

\begin{document}
\begin{flushright}
SISSA-ISAS 53/95/EP\\
UT--704\\
hep-th/9505140
\end{flushright}
\vskip0.5cm
\centerline{\LARGE\bf Topological field theory interpretations}
\vskip0.2cm
\centerline{\LARGE \bf and LG representation of $c=1$ string theory}
\vskip1cm
\centerline{\large  L.Bonora\footnote{ E-mail: bonora@tsmi19.sissa.it} }
\centerline{International School for Advanced Studies (SISSA/ISAS)}
\centerline{Via Beirut 2, 34014 Trieste, Italy}
\centerline{INFN, Sezione di Trieste.  }
\vskip0.5cm
\centerline{\large C.S.Xiong\footnote{E-mail:
xiong@danjuro.phys.s.u-tokyo.ac.jp}}
\centerline{Department of Physics, University of Tokyo}
\centerline{Bunkyo-ku 7-3-1, Tokyo, Japan}
\vskip3cm
\abstract{We analyze the topological nature of $c=1$ string theory at the
self--dual radius. We find that it admits two distinct topological field theory
structures characterized by two different puncture operators.
We show it first in the unperturbed theory in which the only parameter
is the cosmological constant, then in the presence of any infinitesimal
tachyonic perturbation. We also discuss in detail a Landau--Ginzburg
representation of one of the two topological field theory structures.}
\vfill\eject

\section{Introduction}

The topological nature of $c=1$ string theory, \cite{c=1},
at the self--dual radius
has been recently studied by several authors. The main tool in this regard
has been provided by the remark that the structure underlying this theory
is the $2$--dimensional dispersionless Toda lattice hierarchy,
whose two series of flow parameters play the role of coupling constants
of the purely tachyonic states $T_n$ and $T_{-n}$.
In refs.\cite{GMHO}, it has been proposed that $T_1$
play the role of {\it puncture operator} and
the tachyons be primary fields.

In \cite{BX1},\cite{BX2} we have shown that the two--matrix model
provides a good description of $c=1$ string theory.
In fact: 1) two--matrix model has allowed us
to reproduce all the known results of $c=1$ string theory
(which is not surprising
since it is equivalent to the Toda lattice hierarchy constrained
with suitable boundary conditions); 2) it has allowed us to introduce
additional discrete states, which can be interpreted as the discrete states
of $c=1$ string theory, and to compute their correlation functions;
3) two--matrix model is defined for every genus.

The present paper is specifically devoted to the discussion of the
topological nature of
$c=1$ string theory. In \cite{BX2} we suggested that there may exist another
topological field theory (TFT) interpretation in which the puncture
operator is $T_0$,
the primary fields are all the pure tachyonic states, while the descendants
are the discrete states.  Here we show that this interpretation as well as
the previous one, \cite{GMHO}, with some adjustments, are both legitimate.
We prove this in the pure cosmological sector of the theory (in which the
only parameter is the cosmological constant), but we substantiate our
assertion by showing that the axioms of topological field theory hold in
the presence of any (infinitesimal) perturbation by the tachyonic states
and by providing a Landau--Ginzburg description of the structure which has
$T_1$ as puncture operator. Actually there is a third topological field theory
structure in which $T_{-1}$ is the puncture operator, but this structure is
exactly specular to the latter.

With the evidence we provide in this paper we think it is justified
to say that $c=1$ string theory at the self--dual radius (and two--matrix
model with it) is a huge topological field theory which is defined at
all genera. In the final section we suggests that other smaller TFT's are
imbedded in such huge theory, and this perhaps provides a clue to understanding
why $T_0$ and $T_1$ (or $T_{-1}$) can be both interpreted as puncture
operators.

We remark that so far only TFT's with a finite number of primaries have been
fully analyzed and coupled to topological gravity, \cite{WDVV},
-- we have in mind typically the ADE models, let us generically refer to
them as $c<1$ TFT's. The TFT's we are
considering here are of a quite different type: first of all they have an
infinite number of primaries, secondly the states are classified according
to $sl_2$ representations (instead of a $U(1)$ charge label). It is therefore
non--trivial not only that they satisfy the TFT axioms, but that the coupling
to topological gravity occurs with essentially the same rules as in the
$c<1$ models. It is remarkable that all these properties, as well as their
generalization at every genus, are embodied in the
Toda hierarchy subject to the coupling conditions of the two--matrix models.

The paper is organized as follows. Below we summarize some notations and
formulas which will be used throughout the article. In section 2 we introduce
the two (three) topological field theory structures announced above in the
purely cosmological sector and show that all the axioms of TFT's are satisfied.
In section 3 we show the same thing in the presence of any infinitesimal
tachyonic perturbation. In section 4 we summarize the equations characterizing
the dispersionless Toda hierarchy and restrict it to the $c=1$ string theory.
With this material, in section 5, we present a Landau-Ginzburg (LG)
interpretation of the TFT structure in which $T_1$ is the puncture operator.

\subsection{Genus 0 correlators of the discrete states}

We collect in this section some results taken from \cite{BX1},\cite{BX2}
we will need in the following. We start with some notations. We will label the
states
of the theory with latin letters: the first latin letters
$a,b,c,...$ will be used to denote integers, while $i,j,k,l,m,n,...$ will
denote
positive integers and $r,s$ non--negative integers. The tachyonic states will
be
denoted $T_n, T_{-n}$. The cosmological
operator $T_0$ will also be denoted $Q$. So, in particular,
the set $\{T_a\}$ is the same as the set $\{T_n, T_0,T_{-n}\}$.
The generic discrete states
are denoted $\chi_{r,s}$, and we have the correspondence $\chi_{n,0}= T_n$,
$\chi_{0,n}=T_{-n}$ and $\chi_{0,0}=T_0$; the remaining discrete states, with
both $r$ and $s$ non--vanishing, are called extra.

Each discrete state is coupled to the theory via a coupling $g_{r,s}$. We use
the convention
$g_{n,0}\equiv t_{1,n}$ and $g_{0,n}\equiv t_{2,n}$, while $g_{0,0}$ is
identified
with the cosmological constant $x$.
The model or sector of the whole theory that results when we switch on the
couplings $t_{1,1},...,t_{1,p},t_{2,1},...,t_{2,q}$ (beside $g_{0,0}$ and
$g_{1,1}$) will be
called ${\cal M}_{p,q}$. Throughout the paper we set $g_{1,1}=-1$.
Let us denote by ${\cal S}_0$ the model
${\cal M}_{0,0}$ restricted by the condition $g_{1,1}=-1$. ${\cal S}_0$ is
what we mean by purely cosmological sector of the $c=1$ string theory.

The genus 0 correlation functions (CF's), denoted with $<\cdot>$ throughout
the paper, for ${\cal S}_{0}$ are given by
\a
<\chi_{n,n}>= {x^{n+1}\over {n+1}}\label{chinn0}
\b
\a
<\chi_{r_1, s_1} \chi_{r_2,s_2}> = {x^\Sigma }
{{M(r_1,s_1)M(r_2,s_2)}\over \Sigma}\label{chi20}
\b
where $\Sigma= r_1+r_2=s_1+s_2$ and $M(r,s)= max(r,s)$.
This formula also holds when the two labels of $\chi$ coincide.
\a
<\chi_{r_1,s_1}\chi_{r_2,s_2}\chi_{r_3,s_3}> = {x^{\Sigma-1}}
 M(r_1,s_1)M(r_2,s_2)M(r_3,s_3)\label{chi30}
\b
where $\Sigma = r_1+r_2+r_3=s_1+s_2+s_3$.

For the n--point functions with ${\rm n}>3$, as is by now well--known,
there is more than one possibility. We give here only
\a
&&<\chi_{r_1,s_1}\ldots \chi_{r_n,s_n}> =
{x^{\Sigma -n+2}}
M(r_1,s_1)\dots M(r_n,s_n)(\Sigma -1)\ldots(\Sigma -n +3)\0\\
&& \Sigma = r_1+\ldots+r_n = s_1 +\ldots +s_n \label{chin0}
\b
if $\Sigma >n-2$, and vanishes otherwise. This formula holds when
there is one label $r_k>s_k$ and $n-1$ labels $r_k<s_k$.

We remark that the above formulas have been derived for states $\chi_{r,s}$
with $r$ and $s$ not simultaneously vanishing. To obtain CF's involving $p$
insertions of $Q\equiv \chi_{0,0}$, one has simply to differentiate
$p$ times with respect to $x$ the corresponding CF without $Q$ insertions.
For CF's containing only $Q$ insertions, we have
\a
F_h= \chi_h^{(0)} x^{2-2h},
\qquad
<Q^n>_h = n! \chi_{h}^{(n)} x^{2-2h-n}\label{Qn}
\b
where
\a
\chi_{h}^{(n)}= \frac{(-1)^n(2h-3+n)!(2h-1)} {n!(2h)!}B_{2h}\0
\b
Here the label $h$ denotes the genus $h$ contribution. $\chi_{h}^{(n)}$
is the virtual Euler characteristic of the moduli space of Riemann surfaces
of genus $h$, \cite{HZP}. We see therefore that $Q$, the operator
coupled to $x$, is
to be interpreted in this context as the puncture operator.

\section{TFT interpretations}

\subsection{First interpretation: puncture operator $T_1$}.

First we recall that a (matter) topological field theory is defined by a
(usually finite) set of primary fields $\phi_\al$, $\al = 1,2,...$.
Among them one, say $\phi_1$, plays a special role.
The n--th point correlators
are pure numbers and, in particular, the 3--point functions (in genus 0)
$C_{\al,\beta,\gamma}=<\phi_\al\phi_\beta\phi_\gamma>$ are crucial in the
definition of TFT. The metric $\eta_{\al,\beta}$ coincides by definition
with $C_{1,\al,\beta}$ and is required to be invertible. The inverse metric
is denoted by $\eta^{\al,\beta}$. The last defining property of TFT is the
associativity condition
\a
\sum_{\lambda ,\mu}C_{\al,\beta,\lambda}\eta^{\lambda,\mu}C_{\mu,\gamma,\delta}
=
\sum_{\lambda ,\mu}C_{\al,\gamma,\lambda}\eta^{\lambda,\mu}C_{\mu,\beta,\delta}
\label{asso}
\b
The coupling of such a theory to topological gravity gives rise to the
descendants and is regulated by two types
of equations, the puncture equations and the recursion relations. $\phi_1$
plays the role of puncture operator.

In the first interpretation of ${\cal S}_0$ as a topological field theory
$T_1$ is the puncture operator and
$T_n \equiv \chi_{n,0}$ and $T_{-n}\equiv \chi_{0,n}$ are the primary
fields, while all the extra states are gravitational descendants.
To justify this assertion we have to find the metric and the structure
constants of the TFT and prove that they satisfy all the axioms.
Moreover we have to define suitable puncture equations and
recursion relations

The metric and the structure constants are given by
\a
\eta_{a.b} = <T_1T_aT_b>,\qquad C_{a,b,c}= <T_aT_bT_c> \label{metric1}
\b
where $a$ and $b$ are integers. The only nonzero elements are
\a
\eta_{n,-n-1}=\eta_{-n-1,n}=<T_1T_nT_{-n-1}> = n(n+1) x^{n},
\quad \eta_{0,1}=\eta_{0,-1} =1\0
\b
This metric is non--degenerate, the inverse is $\eta^{a,b}$ with
\a
\eta^{n,-n-1} = \eta^{-n-1,n} = {x^{-n} \over {n(n+1)}},
\qquad \eta^{1,0} =\eta^{0,-1}=1 \0
\b
while all the other elements vanish. The associativity condition (\ref{asso})
for
the structure constants $C_{a,b,c}$
is easily seen to be satisfied once we notice that the only nonvanishing
three--point functions among primaries are
\a
&&C_{n,m,-n-m} =C_{-n,-m,n+m}
= <T_{-n}T_{-m} T_{n+m}> = nm (n+m)
x^{n+m-1}\0\\
&&C_{n,-m, m-n} = \left\{ \bac nm(n-m)x^{n-1},\quad n>m\\
                               nm(m-n)x^{m-1},\quad n<m\ea\right.\\
&& C_{0,n,m} \equiv n^2 x^{n-1} \delta_{n+m,0},\qquad C_{0,0,0}=x^{-1}\0
\b
The primary fields
form the commutative associative algebra ${\cal A}_1$
\a
T_a T_b = \sum_c C_{a,b}{}^c T_c\label{A1}
\b
where
\a
 C_{a,b}{}^c \equiv
\sum_d C_{a,b,d}\eta^{d,c},\qquad\qquad T_0\equiv Q\0
\b
and we have identified $T_1$ with the identity of ${\cal A}_1$\footnote{In
order
to verify (\ref{A1}) in the correlators one must remember to restore $T_1$ in
the RHS of (\ref{A1})}. To prove ${\cal A}_1$ one has to use
\a
&&C_{n,m}{}^{n+m-1} = \frac {nm}{n+m-1},\qquad C_{0,n}{}^{n-1} ={n\over {n-1}},
\quad C_{n,-n}{}^{-1} = n^2 x^{n-1}, \qquad C_{0,0}{}^{-1} =x^{-1} \0\\
&& C_{n,-m}{}^{n-m-1}= \left\{ \bac \frac{nm}{n-m-1} x^m, \quad n>m+1\\
                               \frac {nm}{m-n+1} x^{n-1}, \quad n<m\ea\right.\\
&&C_{n,-n+1}{}^0 = n (n-1)x^{n-1}\0
\b
where $n,m\neq 0$, $C_{a,b}{}^c =C_{b,a}{}^c$, and the other structure
constants
vanish.

Therefore the set of fields $\{T_{-n} , Q, T_n\}$ define a TFT with puncture
operator $T_1$.

Next let us switch on the coupling to topological gravity, and see whether
this TFT fits into the scheme of a TFT coupled to topological gravity
\cite{WDVV}. We will see that this is indeed the case by exhibiting the
appropriate puncture equations and recursion relations.
For the {\it puncture equations} we start from the string
equation, \cite{BX2},
\a
{\cal L}^{[1]}_{-1}(2)Z(t;g,x)=-T^{[1]}_{-1}(2)Z(t;g,x)\0
\b
where
\a
T_{-1}^{[1]}(2)= \sum_{\stackrel {i\geq 1}
{j\geq 1}}jg_{i,j} \frac{\d}{\d g_{i,j-1} }\0
\b
and differentiate it with respect to $g_{i_1,j_1},\ldots, g_{i_n,j_n}$. Finally
we evaluate it in ${\cal S}_0$. We find
\a
<T_1\chi_{i_1,j_1}\ldots \chi_{i_n,j_n}> = \sum_{l=1}^n j_l
<\chi_{i_1,j_1}\ldots\chi_{i_l,j_l-1}\ldots\chi_{i_n,j_n}>\label{puncture1}
\b
This relation is exact, i.e. valid for all genera.
This holds provided in the LHS there does not appear the operator $T_{-1}$, or,
which is the same, in the RHS there does not appear $Q$. This means the
following: {\it the fields $\chi_{n,m}$, $n,m>0$, are the descendants
of $\chi_{n,0}= T_n$; $Q$ has no descendants; $T_{-n}$ is both a primary and a
descendant of $T_{-1}$.}

Let us pass now to the {\it recursion relations}. They are (in genus 0) and
in ${\cal S}_0$
\a
<\chi_{r,s}\chi_{r_1,s_1}\chi_{r_2,s_2}> = M(r,s) \sum_{l,k}<\chi_{r, s-1}
T_a> \eta^{a,b} <T_b\chi_{r_1,s_1}\chi_{r_2,s_2}>\label{recrel1}
\b
where the labels $k$ and $l$ are understood to be integers.
The proof is very simple. Suppose for example that $r\geq s+1$.
Then
\a
{\rm LHS} = r M(r_1,s_1)M(r_2,s_2) x^{r+r_1+r_2-1}\0
\b
when $r+r_1+r_2 = s+s_1+s_2$ and vanishes otherwise. On the other hand
\a
{\rm RHS} = r <\chi_{r,s-1} T_{s-r+1}> \eta^{s-r+1,r-s}<T_{r-s}
\chi_{r_1, s_1}\chi_{r_2,s_2}> = r M(r_1,s_1)M(r_2,s_2) x^{r+r_1+r_2-1}\0
\b
when $r+r_1+r_2 = s+s_1+s_2$, and vanishes otherwise. The same can be proven
for $r\leq s+1$ \footnote{Eq.(\ref{recrel1}) is true if we exclude
the exceptional case $\chi_{r,s}\neq T_{-1}$,
because we would have in the RHS $<QQ>=\ln x$}.

We can conclude that we have indeed to do with an
unperturbed topological field
theory coupled to topological gravity. We call it ${\cal T}_1$.

The correlators of $c=1$ string theory were obtained in \cite{BX1},
\cite{BX2}, starting from the $W$ constraints of the two-matrix model
(or, equivalently, from the Toda flow equations plus the coupling conditions).
We can therefore say that all we have seen in this section (including the
recursion
relations and the puncture equations) is nothing but a manifestation
of the $W$ constraints. There is also a direct way to prove this assertion,
as pointed out in ref.\cite{BCX}. We have already noticed that the puncture
equations are valid at all genera. As for the other objects and equations
studied in this section, {\it the natural generalization to higher genus is
provided by the $W$ constraints, which are all--genus relations}.

{\bf Remark.} The $<Q^n>$ correlators are exceptional. They are
 not determined by the above
recursion relation and puncture equations, but by the Toda
equation (5.2) of \cite{BX2} which are the appropriate recursion relation for
this kind
of correlators.

Although the Landau--Ginzburg formalism will be discussud at length later on,
let us complete the presentation of ${\cal T}_1$ by anticipating how the
algebra of the primary fields is reproduced in the LG formalism.

The primary fields are represented by polynomials in the variable $\zeta$
and $\zeta^{-1}$
\a
\phi_n \equiv \phi_{n,0} = n \zeta^{n-1}, \qquad
 \phi_0 \equiv \phi_{0,0} =\zeta^{-1},\qquad
 \phi_{-n} \equiv \phi_{0,n} = n x^n \zeta^{-n-1},\label{LGrep2}
\b
These objects form a commutative and associative algebra ${\cal R}_1$ by
simple multiplication. We define the following map
\a
\phi_n \leftrightarrow T_n ,\qquad \phi_0 \leftrightarrow Q,\qquad
\phi_{-n} \leftrightarrow T_{-n}\0
\b
and claim that it is an isomorphism between the algebra ${\cal R}_1$ and the
field algebra ${\cal A}_1$.
It is elementary to prove the isomorphism by checking the few
non--trivial cases.

The TFT interpretation in which the puncture operator is $T_{-1}$,
instead of $T_1$, is perfectly specular (due to the ${\bf Z_2}$
symmetry of the underlying Toda lattice hierarchy under the exchange
of the left with the right sector)
and there is no need to describe it in detail here.
We call the corresponding TFT coupled to topological gravity ${\cal T}_{-1}$.
The physical nature of  the symmetry between left and right sector of
the two-matrix model is not clear. It might be related to some duality
symmetry of the underlying string theory.

\subsection{Another TFT interpretation: puncture operator $Q$.}

In this interpretation the puncture operator is $Q\equiv T_0$. This is
motivated by the fact that, according to the Penner model (see section 1.1),
$Q$ represents a puncture on a Riemann surface. A priori however there is
no compelling reason why this interpretation should work
as the previous one, based on the analogy with the $c<1$
TFT models. The Penner model concerns the virtual Euler
characteristic and involves only correlators of $Q$.
Nevertheless the interpretation turns out to work.

Actually this interpretation has been already introduced in ref.\cite{BX2}.
For the sake of completeness, we review it here and whenever necessary, we
complete
the description given there. The set of primary fields is the same as in
${\cal T}_0$, i.e. $T_n \equiv \chi_{n,0}$, $T_{-n}\equiv \chi_{0,n}$,
and $T_0 \equiv Q$, while all the other $\chi_{n,m}$ are descendants.

The metric is given by
\a
\eta_{a,b} = <QT_aT_b> \label{metric}
\b
where $a$ and $b$ are integers. The only nonzero elements are
\a
\eta_{n,-n}=\eta_{-n,n}=<QT_nT_{-n}> \equiv {\d \over {\d x}}<\chi_{n,0}
\chi_{0,n}> = n^2 x^{n-1}, \quad \eta_{0,0}= x^{-1}\0
\b
This metric is non--degenerate, the inverse is $\eta^{k,l}$ with
\a
\eta^{n,-n} = \eta^{-n,n} = n^{-2} x^{-n+1},\qquad \eta^{0,0} =x \0
\b
while all the other elements vanish. The associativity condition is easily
seen to be satisfied since the only nonvanishing
three--point functions among primaries are
\a
&&C_{n,m,-n-m} =C_{-n,-m,n+m}
= <T_{-n}T_{-m} T_{n+m}> = nm (n+m)
x^{n+m-1}\label{Cijk}\\
&&C_{n,-m, m-n} = \left\{ \bac nm(n-m)x^{n-1},\quad n>m\\
                               nm(m-n)x^{m-1},\quad n<m\ea\right.\0
\b
beside $C_{0,n,m} \equiv \eta_{n,-n} \delta_{n+m,0}$. As is easy to prove,
the primary fields form the commutative associative algebra ${\cal A}_0$
\a
T_a T_b = \sum_c C_{a,b}{}^c T_c,\qquad C_{a,b}{}^c \equiv
\sum_d C_{a,b,d}\eta^{d,c}\0
\b
where again $T_0$ is identified with the identity in ${\cal A}_0$.

The {\it recursion relations} in ${\cal S}_0$ are
\a
<\chi_{r,s}\chi_{r_1,s_1}\chi_{r_2,s_2}> =
M(r,s) \sum_{l,k}<\chi_{r-1, s-1}
T_a> \eta^{a,b} <T_b \chi_{r_1,s_1}\chi_{r_2,s_2}>\label{recrel}
\b
 The proof is very simple and can be found in \cite{BX2}.
\footnote{Here again the relation (\ref{recrel}) does not work when $r=s=1$.
This is an exceptional case due to the fact that on the RHS there appears
the correlator $<QQ>= \ln x$.}

The {\it puncture equations} are designed to connect the CF's of
of the type
\a
<Q \chi_{r_1,s_1} \chi_{r_2 ,s_2} \ldots \chi_{r_n,s_n}>,\0
\b
where the $\chi$'s are extra states,
with CF's including neighboring ascendants of them.
For dimensional reason the latter can only be
$<\chi_{r_1,s_1} \ldots\chi_{r_i-1 ,s_i-1} \ldots \chi_{r_n,s_n}>$.
For the CF's (\ref{chin0}) we have
\a
&&<Q \chi_{r_1,s_1} \chi_{r_2 ,s_2} \ldots \chi_{r_n,s_n}> =
\label{puncture}\\
&&~~~~~~~~~~~~~~~~~~~~~~~
\sum_{i=1}^n\frac{M(r_i,s_i)}{M(r_i-1,s_i-1)}\frac{\Sigma-1}{n}
<\chi_{r_1,s_1} \ldots\chi_{r_i-1 ,s_i-1} \ldots \chi_{r_n,s_n}>\0
\b
where $\Sigma = r_1+\ldots+r_n= s_1 +\ldots +s_n$. In fact the LHS is
\a
<Q\chi_{r_1,s_1} \chi_{r_2 ,s_2} \ldots \chi_{r_n,s_n}>=
x^{\Sigma -n +1}M(r_1,s_1)\ldots
M(r_n,s_n) (\Sigma-1)\ldots (\Sigma -n +2)\0
\b
On the other hand the
generic term in the RHS of (\ref{puncture}) contains
\a
<\chi_{r_1,s_1}\ldots \chi_{r_k-1,s_k-1}\ldots \chi_{r_n ,s_n}> &=&
M(r_1,s_1)\ldots M(r_k-1, s_k-1)\ldots M(r_n,s_n)\cdot \0\\
&&~~~~\cdot(\Sigma-2)\ldots(\Sigma -n+2)x^{\Sigma-n+1}\0
\b
Summing all the contributions in the RHS of (\ref{puncture})
we obtain the equality with the LHS. However a relation similar to
(\ref{puncture}) holds in general. In fact in genus zero the LHS of
(\ref{puncture}) is nothing but the derivative of
\a
<\chi_{r_1,s_1} \chi_{r_2 ,s_2} \ldots \chi_{r_n,s_n}>\label{cf1n}
\b
with respect to $x$. On the other hand
\a
<\chi_{r_1,s_1} \ldots\chi_{r_i-1 ,s_i-1} \ldots \chi_{r_n,s_n}>\label{cf1n'}
\b
is also the derivative of (\ref{cf1n}) with respect to $x$ up to a
multiplicative rational factor. Therefore
by taking a suitable combination of all the (\ref{cf1n'}), we can certainly
reproduce the LHS of (\ref{puncture}).

In summary, in this TFT interpretation {\it
$Q\equiv T_0$ is the puncture operator, $T_n$ and $T_{-n}$ ($n$ positive)
are the primary fields, while $\chi_{n+k,k}$ and $\chi_{k,n+k}$, with $k$
positive, are, respectively, the descendants. In particular $\chi_{k,k}$
are the descendants of $Q$. Once again we have to do with an unperturbed TFT
coupled to
topological gravity. Let us call it ${\cal T}_0$.}

It is evident that the puncture equation is determined by the
dispersionless flow in $x$, i.e.
in $N$, the size of the matrices in the two--matrix model. Therefore
it does not extend, as it is, to higher genus.
For example, for one point functions, the all--genus puncture equation becomes
\a
<(1- e^{-Q})\chi_{r,r}>_{\rm all-genus} = r<\chi_{r-1,r-1}>_{\rm
all-genus}\label{punctureallg}
\b
This is exact and is clearly the generalization of (\ref{puncture}) to
every genus.

{\it The rule is very simple: to generalize (\ref{puncture}) one has
simply to write down the exact flow in N}. The latter is provided by the
two--matrix model. In general we can repeat the same
conclusion as in the previous subsection:
{\it the higher genus puncture and recursion relations are nothing but
the $W$ constraints of the two matrix model}.

For the $<Q^n>$ correlators the same remarks holds as in the previous
subsection.

We showed in \cite{BX2} that a LG interpretation of ${\cal T}_0$
can be introduced, but we will not insist here on such interpretation.

\section{Perturbation of the TFT's.}

We study now the TFT's ${\cal T}_0$ and ${\cal T}_1$ under the most general
infinitesimal tachyonic perturbation. Since in the infinitesimal case,
the perturbation by an operator $T_n$ or $T_{-n}$ appears linearly, it is
enough to consider such perturbations one
by one. Let us study hereby the perturbations by $T_p$ and by $T_{-q}$,
where $p$ and $q$ are positive integers.

The relevant correlators (one, two and three--point functions) perturbed by
means of $T_1$ and $T_{-1}$ have been given in \cite{BX2}. The ones perturbed
by means of $T_2$ and $T_{-2}$ have been given in \cite{BCX}. Finally the
correlators infinitesimally perturbed by $T_p$ and $T_{-q}$ can be obtained
from the formulas of the following sections or by solving the coupling
conditions in the two matrix model.

Below we give the results relevant to the subsequent developments.
Let us set
\a
p t_{1,p}= \epsilon_p\ll 1,\qquad\quad qt_{2,q}= \zeta_q\ll 1\0
\b
Then we have
\a
C_{n,m,l} &=& nml(q-1)\zeta_q x^{q-2} \delta_{n+m+l,q}\0\\
C_{n,m,-l} &=& nm(n+m)x^{n+m-1} \delta_{l,n+m} +\ep_p nml(l-1) x^{l-2}
\delta_{l,n+m+p}+\0\\
&&\zeta_q nml\Big( (l-1)\th(l-m)\th(l-n) + (n-1)\th(n-l)\th(l-m)+
\label{Cpert1}\\
&&(m-1)\th(m-l)\th(l-n) +
(q-1)\th(n-l)\th(m-l)\Big) x^{l+q-2}\delta_{l+q,n+m}\0\\
C_{n,-m,-l} &=& nmlx^{n-1}\delta_{n,m+l} + \ep_p nml\Big(
(n-1)\th(n-m)\th(n-l)+
(m-1)\th(m-n)\th(n-l) +\0\\
&& (l-1)\th(n-m)\th(l-n) + (p-1)\th(l-n)\th(m-n)\Big)x^{p+n-2}
\delta_{n+p,m+l}+\0\\
&&\zeta_q nml(n-1)x^{n-2}\delta_{n,m+l+q}\0\\
C_{-n,-m,-l} &=& \ep_p nml (p-1)x^{p-2} \delta_{p,n+m+l}\0
\b
We have moreover
\a
C_{0,n,m} &=& \zeta_q nm(q-1) x^{q-2} \delta_{q,n+m}\0\\
C_{0,n,-m} &=& n^2x^{n-1}\delta_{n,m} + \ep_p nm(m-1)x^{m-2}\delta_{m,n+p}+
\zeta_q nm(n-1) x^{n-2} \delta_{m+q,n}\0\\
C_{0,-n,-m} &=& \ep_p nm(p-1) x^{p-2} \delta_{p,n+m}\label{Cpert2}\\
C_{0,0,n} &=& \zeta_q q(q-1)x^{q-2}\delta_{n,q}\0\\
C_{0,0,-n} &=& \ep_p p(p-1)x^{p-2} \delta_{n,p}\0\\
C_{0,0,0}&=& x^{-1}\0
\b
where
\a
\th(n) =\left\{ \bacc 1, & n>0\\ 1/2, &n=0\\ 0, & n<0\ea\right. \0
\b
As a consequence of the perturbation the metrics and structure constants of
the TFT's are modified. We are going to see next that all the
the topological field theory properties are nevertheless satisfied.

\subsection{$T_1$ puncture operator: perturbation}

In order to show that the $c=1$ string theory with puncture operator $T_1$
perturbed as above is a TFT, we will prove that the inverse metric exists
and that the associativity conditions are satisfied. We prove everything
to the first order in $\ep_p$ and $\zeta_q$, which are infinitesimal.
The metric is given by $\eta_{a,b}= <T_1T_aT_b>$. Its non--vanishing elements
are therefore
\a
\eta_{n,-n-1} &=& n(n+1)x^{n}, \qquad \eta_{0,-1}= 1\0\\
\eta_{n,-n-p-1} &=&\ep_p n(n+p)(n+p+1)x^{n+p-1},\0\\
\eta_{-n,n-p-1} &=& \ep_p n(p-1)(p+1-n)x^{p-1}, \qquad
n<p+1\label{pertmetricT1}\\
\eta_{0, -p-1} &=& \ep_p p(p+1) x^{p-1},\qquad
\eta_{0,q-1} = \zeta_q (q-1)^2x^{q-2}\0\\
\eta_{n, -n+q-1} &=& \zeta_q n (n-1)(n-q+1)x^{n-1}, \qquad n>q-1\0\\
\eta_{n,q-n-1} &=& n(q-1)(q-n-1) x^{q-2},\qquad\quad n<q-1\0
\b
plus the ones which are obtained from these via the symmetry
$\eta_{a,b}=\eta_{b,a}$.

The inverse metric exists, its nonvanishing elements are
\a
\eta^{n,-n-1} &=& \frac{x^{-n}}{n(n+1)},\qquad\quad \eta^{0,-1} =1\0\\
\eta^{n,-n+p-1}&=& - \ep_p\frac{x^{-n+p-1}}{n-p+1},\qquad\qquad n>p-1\0\\
\eta^{n,p-n-1}&=& -\ep_p \frac{p-1}{n(p-n-1)}, \qquad
\qquad n<p-1\label{invmetricT1pert}\\
\eta^{0,p-1} &=& -\ep_p ,\qquad
\eta^{0,-q-1} = -\zeta_q \frac{q-1}{q+1} \0\\
\eta^{n,-n-q-1}&=& -\zeta_q \frac{n+q-1}{n(n+q+1)}x^{-1-n}\0\\
\eta^{-n,n-q-1} &=& -\zeta_q \frac{q-1}{n(q-n+1)}x^{-1},\qquad\qquad n<q+1\0
\b
plus the ones that can be obtained from them via the symmetry
$\eta^{a,b}=\eta^{b,a}$.

One way to prove the associativity conditions for a perturbation
$\ep_p$ is to verify that
\a
C_{n,m,a}\eta^{a,b}C_{b,k,-l} &=&C_{n,k,a}\eta^{a,b}C_{b,m,-l}\0\\
C_{-n,-m,a}\eta^{a,b}C_{b,-k,l} &=&C_{-n,-k,a}\eta^{a,b}C_{b,-m,l}\0\\
C_{n,m,a}\eta^{a,b}C_{b,-k,-l} &=&C_{n,-k,a}\eta^{a,b}C_{b,m,-l}\0\\
C_{-n,-m,a} \eta^{a,b} C_{b,-k,-l} &=& C_{-n,-k,a}\eta^{a,b}C_{b,-m,-l}\0
\b
are identities up to the first order in $\ep_p$. This is a lengthy but
straightforward exercise on the basis of (\ref{Cpert1},\ref{Cpert2},
\ref{invmetricQpert}). These four identities are enough since all the other
identities that appear in (\ref{asso}) can be obtained  from them either
by using the symmetry properties of $C$ and $\eta$, or, when some of the
indices
$n,m,k,l$ are replaced by 0, by remarking that we have formally
\a
C_{0,a,b} = {\rm lim}_{n\rightarrow 0}\frac{C_{n,a,b}}{n},\qquad\quad
C_{0,0,b} = {\rm lim}_{n,m\rightarrow0}\frac{C_{n,m,b}}{nm}\0
\b
We proceed in a similar way with a $\zeta_q$ perturbation.

The metric (\ref{pertmetricT1}) depends in general on the
perturbation parameters,
but it is easy to find redefinitions of the primaries so as to
recover a constant metric. For example, for a $\ep_p$ perturbation, we
define
\a
&&\hat T_n = T_n - \ep_p\frac {n(n+\frac{p-1}{2})}{n+p} x^{-1}T_{n+p}\0\\
&&\hat T_{-n} = T_{-n} - \ep_p {1 \over 4}\frac{n(p-3)}{p-n} x^{n-1} T_{p-n},
\qquad n<p\0\\
&&\hat T_{-p} = T_{-p} -\ep_p {1 \over 4}p(p-3) x^{p-1}T_0\label{redefT1}\\
&&\hat T_{-n} = T_{-n}, \qquad n> p\0\\
&&\hat T_0 = T_0 -\ep_p \frac {x^{-1}}{2} \frac{p-1}{p} T_{p}\0
\b
In terms of the hatted fields the metric becomes constant and equal to the
unperturbed one. A similar redefinition can be done for any $\zeta_q$
perturbation.

As for the coupling to topological gravity, the puncture equations and
recursion
relations have to be, in general, suitably modified with respect to the
previous section. There is however no point in writing them down explicitly.
They are nothing but particular aspects of the $W$--constraints of the
two--matrix model.

The situation with $T_{-1}$ as puncture operator is exactly specular.

\subsection{Q puncture operator: perturbation}

We now do the same when $Q$ is the puncture operator.
The metric is given by $\eta_{a,b}=$ $<QT_aT_b>$. The only non--vanishing
metric elements are therefore
\a
\eta_{n,-n} &=& n^2x^{n-1}, \qquad \eta_{0,0}= x^{-1}\0\\
\eta_{n,-n-p} &=&\ep_p n(n+p)(n+p-1)x^{n+p-2},\0\\
\eta_{-n,n-p} &=& \ep_p n(p-1)(p-n)x^{p-2}, \qquad n<p\label{pertmetricQ}\\
\eta_{0, -p} &=& \ep_p p(p-1) x^{p-2},\qquad
\eta_{0,q} = \zeta_q q(q-1)x^{q-2}\0\\
\eta_{n, -n+q} &=& \zeta_q n (n-1)(n-q)x^{n-2}, \qquad n>q\0\\
\eta_{n,q-n} &=& \zeta_q n(q-1)(q-n) x^{q-2},\qquad n<q\0
\b
plus the ones which are obtained from these via the symmetry
$\eta_{a,b}=\eta_{b,a}$.

The inverse metric exists, its nonvanishing elements are
\a
\eta^{n,-n} &=& \frac{x^{1-n}}{n^2},\qquad\quad \eta^{0,0} =x\0\\
\eta^{n,-n+p}&=& - \ep_p\frac{n-1}{n(n-p)} x^{-n+p},\qquad\qquad n>p\0\\
\eta^{n,p-n}&=& -\ep_p \frac{p-1}{n(p-n)},\qquad \qquad
n<p\label{invmetricQpert}\\
\eta^{0,p} &=& -\ep_p \frac{p-1}{p},\qquad
\eta^{0,-q} = -\zeta_q \frac{q-1}{q} \0\\
\eta^{n,-n-q}&=& -\zeta_q \frac{n+q-1}{n(n+q)}x^{-n}\0\\
\eta^{-n,n-q} &=& -\zeta_q \frac{q-1}{n(q-n)},\qquad\qquad n<q\0
\b
plus the ones that can be obtained from them via the symmetry
$\eta^{a,b}=\eta^{b,a}$.

Using these formulas one can prove the associativity conditions in the same
way as above.

The metric (\ref{pertmetricQ}) depends in general on the perturbation
parameters.
However it is easy to find redefinitions of the primaries so as to
recover a constant metric. For example, for a $\ep_p$ perturbation, we
define
\a
&&\hat T_n = T_n - \ep_p\frac {n(n+\frac{p-1}{2})}{n+p} x^{-1}T_{n+p}\0\\
&&\hat T_{-n} = T_{-n} - \ep_p {1\over 4}\frac{n(p-1)}{p-n} x^{n-1} T_{p-n},
\qquad n<p\0\\
&&\hat T_{-n} = T_{-n}, \qquad n\geq p\label{redefQ}\\
&&\hat T_0 = T_0 - \ep_p \frac {x^{-1}}{2} \frac{p-1}{p} T_{p}\0
\b
In terms of the hatted fields the metric becomes constant and equal to the
unperturbed one. A similar redefinition can be done for any $\zeta_q$
perturbation.

\section{The $c=1$ string theory and the extended Toda lattice hierarchy}

\setcounter{equation}{0}
\setcounter{subsection}{0}

Let us now turn our attention to the Toda lattice hierarchy.
This allows us on one the hand to compute correlators when finite
perturbations are switched on -- in particular one can derive
the results used in the previous sections.
On the other hand we prepare the ground to introduce, in
the following section, a LG representation of ${\cal T}_1$. We recall
that in the extended (restricted) Toda lattice hierarchy the extra states
are admitted (excluded).

We have already pointed out that
the $c=1$ string theory is described at the self--dual point by the extended
two matrix model, which is equivalent to the extended $2d$ Toda lattice
hierarchy subject to the coupling conditions. In this section we
will only pay attention to the genus zero case, therefore
we will briefly review the dispersionless extended Toda lattice hierarchy,
and its restriction to the $c=1$ string theory.

\subsection{Dispersionless extended Toda lattice hierarchy}

The dispersionless extended Toda hierarchy is based on four objects,
which are Laurent series in the complex variable $\zeta$.
Two of them are the so--called Lax operators
\a
L=\zeta + \sum_{l=0}^\infty a_l \zeta^{-l}, \qquad
\tl=\frac{R}{\zeta} + \sum_{l=0}^\infty \frac{b_l}{R^l} \zeta^l, \label{ltl}
\b
The other two are
\a
&& M= \sum_{r=1}^\infty rt_{1,r} L^{r-1} + x L^{-1} +
 \sum_{r=1}^\infty {{\d F}\over {\d t_{1,r}}}L^{-r-1}, \label{cm}\\
&&\hs({\widetilde M}) = \sum_{r=1}^\infty rt_{2,r} {\widetilde L}^{r-1} + x
{\widetilde L}^{-1} +
\sum_{r=1}^\infty {{\d F}\over {\d t_{2,r}}}{\widetilde
L}^{-r-1}. \label{ctm}
\b
where the operation $\hs$ is defined as follows
\a
\hs(\zeta)=\frac{R}{\zeta}, \qquad \hs(f)=f, \quad \forall~{\rm function}
{}~~f. \label{sigma}
\b
In these equations, $F$ is connected with the $\tau$--function (see below) and
will be subsequently interpreted as the free energy.
$R,a_l$ and $b_l$ are `fields' (i.e. functions of the couplings).
With respect to the basic Poisson bracket
\a
\{\zeta, ~~x\} = \zeta, \label{zetax}
\b
the four Laurent series given above satisfy the fundamental relations
\a
\{ L, ~~~ M\}=1, \qquad \{ \hs(\tl), ~~~ \tm\} = 1. \label{lmtltm}
\b
The dispersionless extended Toda hierarchy can be represented
in several different ways.

\bigskip
\noindent
\underline{\it The first representation} :

The dispersionless extended Toda hierarchy
can be written as follows
\ai
&&{{\d L}\over {\d g_{r,s}}} = \{ L, ~~ (L^r\tl^s)_-\}, \label{distoda1}\\
&&{{\d \tl}\over {\d g_{r,s}}} = \{ (L^r\tl^s)_+, ~~\tl\}, \label{distoda2}
\bj
where $g_{r,s}$ are the flow parameters or coupling constants
with non--negative integers  $(r,s)$ (they are not simutaneously zero)
introduced in section 1. For any Laurent
series $f(\zeta)=\sum_{i}f_i\zeta^i$, we denote
\a
f_+(\zeta)=\sum_{i\geq0}f_i\zeta^i, \qquad
f_-(\zeta)=\sum_{i<0}f_i\zeta^i,\0
\b
and
\a
f_{\geq k}(\zeta)=\sum_{i\geq k}f_i\zeta^i, \qquad
f_{\leq l}(\zeta)=\sum_{i\leq l}f_i\zeta^i, \qquad
f_{(k)}(\zeta)=f_k.\0
\b
For the sake of completeness, one may add to
eqs.(\ref{distoda1},\ref{distoda2}) two $x$--flow equations
\a
{{\d L}\over {\d x}} = \{ (\ln L)_+, ~~ L\}, \qquad
{{\d \tl}\over {\d x}} = \{ (\ln\tl)_+, ~~\tl\}. \label{0flow}
\b

The {\it $\tau$-function}\, of the above integrable hierarchy (denoted
by $e^F$) is linked to the fields $a_l$ and $b_l$ thorough the
following relation
\a
\frac{\d}{\d g_{r,s}}F = \int_0^x\, (L^r\tl^s)_{(0)}(y)dy. \label{ddf}
\b
This relation, together with the flow equations, leads to
\a
\frac{\d^2}{\d t_{1,1} \d t_{2,1}}\ln R = \d^2_x R, \label{contodaeq}
\b
which is the continuum version of the $2d$ Toda lattice equation.
One may also  re--express the dispersionless extended
Toda hierarchy in terms of $M$ and $\tm$
\ai
&&{{\d M}\over {\d g_{r,0}}} = \{ L^r_+, ~~M\}, \qquad r\geq1 \label{ddgm1}\\
&&{{\d M}\over {\d g_{r,s}}} = \{ M, ~~ (L^r\tl^s)_-\}, \qquad r\geq0,~~s\geq1
\label{ddgm2}\\
&&{{\d \hs(\tm)}\over {\d g_{0,s}}} = \{ \hs(\tm), ~~\tl^s_-\},
  \qquad s\geq1 \label{ddgm3}\\
&&{{\d \hs(\tm)}\over {\d g_{r,s}}} = \{ (L^r\tl^s)_+, ~~\hs(\tm)\},
  \qquad r\geq1,~~s\geq0. \label{ddgm4}
\bj
Both representations (\ref{distoda1},\ref{distoda2}) and
(\ref{ddgm1}--\ref{ddgm4}) will be useful in our later discussion.

It is useful to express the Toda lattice in terms of the underlying
linear systems, i.e. by means of suitable {\it Baker--Akhiezer functions}.
The Baker--Akhiezer functions $\Psi(\lambda_1)$ and $\Psi(\lambda_2)$,
appropriate for our case, are given by
\a
&&\ln\Psi(\lambda_1) = \sum_{r=1}^\infty t_{1,r} \lambda_1^r +
x \ln \lambda_1 - \sum_{r=1}^\infty \frac{1}{r\lambda_1^r}
\frac{\d F}{\d t_{1,r}}, \label{cpsin} \\
&&\ln\Phi(\lambda_2) = \sum_{r=1}^\infty t_{2,r} \lambda_2^r +
x \ln \lambda_2 - \sum_{r=1}^\infty \frac{1}{r\lambda_2^r}
\frac{\d F}{\d t_{2,r}}. \label{cphin}
\b
The spectral parameters $\lm_1,\lm_2$ and the Lax operators $L, \hs(\tl)$
are interchangeable, so we have
\a
M(L) = \frac{d \ln\Psi(L)}{d L}, \qquad \qquad
\tm(\tl) = \frac{d \ln\Phi(\hs(\tl))}{d \hs(\tl)}.  \label{dcuctu}
\b
The equations of motion of Baker--Akhiezer functions are
\ai
&&{{\d \ln\Psi(L)}\over {\d x}} = \ln\zeta, \qquad
{{\d \ln\Phi(\hs(\tl))}\over {\d x}} = \ln\zeta, \label{dgg0}\\
&&{{\d \ln\Psi(L)}\over {\d g_{r,0}}} =  L^r_+(\zeta),
  \qquad r\geq1; \label{dggpsi1}\\
&&{{\d \ln\Psi(L)}\over {\d g_{r,s}}} = - (L^r\tl^s)_-(\zeta),
  \qquad r\geq0,~~s\geq1; \label{ddgpsi2}\\
&&{{\d \ln\Phi(\hs(\tl))}\over {\d g_{0,s}}}
  = \Bigl(\hs(\tl^s) \Bigl)_+(\zeta),
  \qquad s\geq1; \label{dggphi1}\\
&&{{\d \ln\Phi(\hs(\tl))}\over {\d g_{r,s}}}
  = -\Bigl(\hs(L^r\tl^s)\Bigl)_-(\zeta),
  \qquad r\geq1,~~s\geq0. \label{ddgphi2}
\bj
where the arguments in the brackets on the LHS are fixed
when taking derivatives with respect to the coupling parameters.
These equations, on one hand reproduce eqs.(\ref{ddgm1}--\ref{ddgm4}),
on the other hand lead (together with eqs.(\ref{cpsin},\ref{cphin})) to
\ai
&& (L^r\tl^s)_- = \sum_{k=1}^\infty \frac{1}{kL^k} \frac{\d^2 F}
 {\d g_{k,0} \d g_{r,s} }, \label{lrtls}\\
&&(L^r\tl^s)_{\geq1} = \sum_{k=1}^\infty \frac{1}{k\tl^k}
 \frac{\d^2 F}{\d g_{0,k} \d g_{r,s} }, \label{lrtls'}
\bj
where $r,s$ are non--negative integers (not simutaneously zero). In the case
$(r,s)=(0,0)$, we have
\ai
&&\ln\zeta = \ln L - \sum_{k=1}^\infty \frac{1}{kL^k}
 \frac{\d^2 F}{\d g_{k,0} \d x}, \\
&&\ln\zeta = \ln \hs(\tl) - \sum_{k=1}^\infty \frac{1}{k\hs(\tl^k)}
 \frac{\d^2 F}{\d g_{0,k} \d x}. \label{lnzeta}
\bj
Combining eqs.(\ref{lrtls}--\ref{lnzeta}), we are able to obtain
the following identities
\a
 L^r\tl^s = \sum_{k=1}^\infty \frac{1}{kL^k} \frac{\d^2 F}
 {\d g_{k,0} \d g_{r,s} } + \frac{\d^2 F}{\d x \d g_{r,s} }
+ \sum_{k=1}^\infty \frac{1}{k\tl^k}
 \frac{\d^2 F}{\d g_{0,k} \d g_{r,s} }, \label{ltlf}
\b
where $(r,s)\neq(0,0)$, and
\a
\ln(L\tl) = \sum_{k=1}^\infty \frac{1}{kL^k} \frac{\d^2 F}
 {\d g_{k,0} \d x } + \ln R
+ \sum_{k=1}^\infty \frac{1}{k\tl^k}
 \frac{\d^2 F}{\d g_{0,k} \d x }. \label{ltlf0}
\b
On the other hand, eq.(\ref{lrtls}) immediately leads to
\a
 \frac{\d^2 F} {\d g_{k,0} \d g_{r,s} } = \oint (L^r\tl^s)_- d L^k,
 \label{h2pf}
\b
similarly
\a
 \frac{\d^2 F}{\d g_{0,k} \d g_{r,s} }=\oint
(L^r\tl^s)_{\geq1}d\tl^k. \label{h2pf'}
\b
We remark that these formulas are valid in general, without any
restriction on the couplings,
and even before trucating to $c=1$ string theory, as will be done in the
following subsection.

\bigskip
\noindent
\underline{\it The second representation} :

Equivalently we may represent the integrable hierarchy in terms of
two conjugate pairs $(L,M)$ and $(\tl, \tm)$ and
infinite many Poisson brackets. From eqs.(\ref{distoda1},
\ref{distoda2}), and eqs.(\ref{ddgm1}--\ref{ddgm4}), as well as
eqs.(\ref{lmtltm}), we can derive
\ai
\frac{d}{d \zeta} \Bigl( L^i(\zeta)\Bigl)_+
&=& \frac{\d M(\zeta)}{\d t_i} \frac{\d L}{\d \zeta}
-\frac{\d L(\zeta)}{\d t_i} \frac{\d M}{\d \zeta}, \label{ddtim}\\
\frac{d}{d \zeta} \Bigl( (L^i\tl^j)(\zeta)\Bigl)_-
&=& \frac{\d L(\zeta)}{\d g_{i,j}} \frac{\d M}{\d \zeta}
-\frac{\d M(\zeta)}{\d g_{i,j}} \frac{\d L}{\d \zeta}, \qquad i\geq0,j\geq1,
\label{ddgijm}\\
\frac{d}{d \zeta} \Bigl( \hs(\tl^j)(\zeta)\Bigl)_+
&=& \frac{\d \tm(\zeta)}{\d \tt_j} \frac{\d \hs(\tl)}{\d \zeta}
-\frac{\d \hs(\tl)(\zeta)}{\d\tt_j}\frac{\d \tm}{\d \zeta}, \label{ddttjtm}\\
\frac{d}{d \zeta} \Bigl(\hs(L^i\tl^j)(\zeta)\Bigl)_-
&=& \frac{\d \hs(\tl)(\zeta)}{\d g_{i,j}} \frac{\d \tm}{\d \zeta}
-\frac{\d \tm(\zeta)}{\d g_{i,j}}\frac{\d\hs(\tl)}{\d \zeta},
 \qquad i\geq1,j\geq0. \label{ddgijtm}
\bj
These equations have a Poisson bracket structure which can be made explicit
by introducing one Poisson bracket for each coupling constant
\a
\{\zeta, ~~ g_{i,j}\}_{i\otimes j} = \zeta. \label{zetagij}
\b
In terms of these Poisson brackets, the above flow equations
can be rewritten as
\ai
\zeta\frac{d}{d \zeta} \Bigl( L^i(\zeta)\Bigl)_+
&=& \{L(\zeta),  ~~~M(\zeta)\}_{i\otimes0}, \label{ddtim'}\\
-\zeta \frac{d}{d \zeta} \Bigl( (L^i\tl^j)(\zeta)\Bigl)_-
&=& \{L(\zeta),  ~~~M(\zeta)\}_{i\otimes j}, \qquad i\geq0,j\geq1,
\label{ddgijm'}\\
\zeta\frac{d}{d \zeta} \Bigl( \hs(\tl^j)(\zeta)\Bigl)_+
&=& \{\hs(\tl)(\zeta),  ~~~\tm(\zeta)\}_{0\otimes j}, \label{ddtitm'}\\
-\zeta \frac{d}{d \zeta} \Bigl( \hs(L^i\tl^j)(\zeta)\Bigl)_-
&=& \{\hs(\tl)(\zeta), ~~~ \tm(\zeta)\}_{i\otimes j}, \qquad i\geq1,j\geq0.
 \label{ddgijtm'}
\bj
This is our second representation of the extended dispersionless
Toda hierarchy. The first description is a good framework
to describe Hamiltonian structures. The second
representation will naturally lead to the Landau--Ginzburg formulation.

Actually there other possible formulations of the hierarchy. For example
after introducing the {\it canonical} momentum $p\equiv \zeta +a_0$, we
can reformulate the hierarchy in terms of the canonical Poisson bracket
\a
\{ p, ~~~t_{1,1}\}_{\rm KP} =1,\0
\b
instead of using (\ref{zetax}). The resulting form of the hierarchy
is nothing but the extension of the standard dispersionless KP hierarchy.
The flows corresponding to discrete states are related to additional
symmetries of the KP hierarchy.

\subsection{Truncation to the $c=1$ string theory}

Let us introduce some notations
\a
<\chi_{i_1,j_1}\chi_{i_2,j_2}\ldots \chi_{i_n,j_n}>
\equiv \frac { \d^n F} { \d g_{i_1,j_1}\d g_{i_2,j_2}\ldots \d g_{i_n,j_n}},
\qquad
<Q>\equiv \frac{\d F}{\d x}. \0
\b
Beside ${\cal S}_0$, we define three more particular subspaces of the
full coupling space
\a
&&{\cal S} = \{ t_{1,r},\qquad t_{2,s}, \qquad \forall r,s\geq1, \qquad
x,\qquad  g_{1,1}=-1 \}; \0\\
&&{\cal S}_+ = \{ t_{1,r}, \qquad \forall r\geq1, \qquad t_{2,1}, \qquad
 x, \qquad g_{1,1}=-1 \}; \0\\
&&{\cal S}_- = \{ t_{1,1}, \qquad t_{2,s}, \qquad \forall s\geq1, \qquad
 x,\qquad  g_{1,1}=-1 \}; \0
\b
The number of couplings in ${\cal S}, {\cal S}_+$ and ${\cal S}_-$
is understood to be arbitrarily large but finite, therefore ${\cal S}$ denotes
any model ${\cal M}_{p,q}$, ${\cal S}_+$ any model ${\cal M}_{p,1}$ and
${\cal S}_-$ any model ${\cal M}_{1,q}$, evaluated at $g_{1,1} =-1$.

The extended Toda lattice hierarchy provides the description of $c=1$ string
theory, {\it if and only if}\, we impose certain constraints, the coupling
conditions. In the
dispersionless limit, the fundamental constraints (or {\it coupling
conditions}) take the following simple form
\a
M + \sum_{r,s\geq1}rg_{r,s}L^{r-1}\tl^s=0, \qquad
\hs(\tm) + \sum_{r,s\geq1} sg_{r,s}L^r \tl^{s-1}=0. \label{discoupling}
\b
After this restriction, the dispersionless $\tau$--function $F$
coincides  with the genus zero free energy of the $c=1$ string theory.
Therefore, $\chi_{i,j}$$(i,j\geq1)$ represent the discrete states, and
$T_i, T_{-j}$ are tachyons, while $Q$ denotes the cosmological operator.

The coupling conditions (\ref{discoupling}) together with the integrable
hierarchy lead to the dispersionless $W_{1+\infty}$ constraints acting
on the free energy. In order to write them down in a compact way,
we define
\a
T_n^{[m]}(1)= \sum_{\stackrel {i_1,\ldots,i_m\geq 1}
{j_1,\ldots,j_m\geq 1}}i_1\ldots i_m g_{i_1,j_1}\ldots
g_{i_m,j_m} \frac{\d}{\d g_{i_1+\ldots+i_m+n-m,j_1+\dots+j_m} },
\quad n\geq0,m\geq1. \label{tnm1}
\b
They satisfy the algebra
\a
[T^{[m]}_n(1), T^{[s]}_r(1)]=(sn-rm)T^{[m+s-1]}_{n+r}(1),
\qquad m,s\geq 1;\quad n,r\geq0. \label{areapres}
\b
This is nothing but the Lie algebra the area--preserving diffeomorphisms.
There is another set of generators
\a
T_n^{[m]}(2)= \sum_{\stackrel {i_1,\ldots,i_m\geq 1}
{j_1,\ldots,j_m\geq 1}}j_1\ldots j_m g_{i_1,j_1}\ldots
g_{i_m,j_m} \frac{\d}{\d g_{i_1+\ldots+i_m,j_1+\dots+j_m+n-m} },
\quad n\geq0,m\geq1. \label{tnm2}
\b
They form another area--preserving diffeomorphism algebra.
In terms of these operators, the constraints can be written as
\a
T^{[m]}_n(1) F = \frac{(-1)^m}{(n+1)(m+1)}\res_\zeta\Bigl(M^{m+1}(L)dL^{n+1}
\Bigl),
\quad m\geq 1,~~n\geq0. \label{wc}
\b
and
\a
T^{[m]}_n(2) F = \frac{(-1)^m}{(n+1)(m+1)}\res_\zeta \Bigl(\hs(M^{m+1})(\tl)
d\tl^{n+1}\Bigl),
\quad m\geq 1,~~n\geq0. \label{wc2}
\b
The simplest case $(n=0, m=1)$ is of particular importance, we may write it
explicitly as follows
\a
 \frac{\d}{\d t_{2,1}}F = \sum_{m,j\geq0, (m,j)\neq(0,0)}
 \Bigl( (m+1)g_{m+1,j}+x t_{1,1} +\delta_{m,0}\delta_{j,1} \Bigl)
\ddg {m,j}F. \label{string2}
\b
Similarly, we have
\a
\frac{\d}{\d t_{1,1}}F = \sum_{i,n\geq0,(i,n)\neq(0,0)}
 \Bigl( (n+1)g_{i,n+1}+x t_{2,1}+\delta_{i,1}\delta_{n,0} \Bigl)
\ddg {i,n}F. \label{string1}
\b
Eq.(\ref{string2}) and eq.(\ref{string1}) are the two simplest constraints,
both of them have the structure of the string equation. So we can choose
either eq.(\ref{string2}), or eq.(\ref{string1}) as string equation.
Thus we have two possible ways to specify the puncture operator,
the primary fields, and the gravitational descendants.  This is
but another manifestation of the duality between the $T_1$ and $T_{-1}$
picture which we have already found in section 3. In order to see this
more closely, let us consider
a general $n$--point function in ${\cal S}_+$,
\a
<\chi_{i_1,j_1}\chi_{i_2,j_2}\cdots \chi_{i_n,j_n}>_{{\cal S}_+}, \0
\b
Differentiating (\ref{string1}) with respect to $g_{i_1,j_1},....,g_{i_n,j_n}$
and evaluating the result in ${\cal S}_+$, we get
\a
<T_1\chi_{i_1,j_1}\chi_{i_2,j_2}\cdots \chi_{i_n,j_n}>_{{\cal S}_+}
=\sum_{l=1}^n j_l
<\chi_{i_1,j_1}\cdots \chi_{i_l,j_l-1}\cdots \chi_{i_n,j_n}>_{{\cal S}_+}\0
\b
This is nothing but the puncture equation (\ref{puncture1}) already found,
extended to the whole ${\cal S}_+$.

Had we done the same thing for (\ref{string2}) in ${\cal S}_-$, we would have
found the analogous puncture equation for ${\cal T}_{-1}$ extended to ${\cal
S}_-$.

All this confirms what we have said before about the identification of the
puncture operators, primary fields and descendants. It is worth remarking that
both puncture equations change their forms if we evaluate them in ${\cal S}$.
In such a case the puncture equation is replaced by just (\ref{string1}) or,
respectively, by (\ref{string2}).

Finally let us have a look at two more constraints.
The next simplest
$W$--constraints correspond the case $n=m=1$. Eq.(\ref{wc}) gives
\a
\ddg {1,1} F &=& \sum_{m,j\geq0}\bigl(mg_{m,j}+\delta_{m1}\delta_{j1}\bigl)
\ddg {m,j} F +\frac{1}{2}x^2,  \label{dilaton2}\\
\ddg {1,1} F &=& \sum_{i,n\geq0}\bigl(ng_{i,n}+\delta_{i1}\delta_{n1}\bigl)
\ddg {i,n} F +\frac{1}{2}x^2.  \label{dilaton1}
\b
As expected, we have two dilaton equations, eq.(\ref{dilaton1}) is
compatible with the string equation (\ref{string1}), while
eq.(\ref{dilaton2}) is the dilaton equation corresponding to
the string equation (\ref{string2}). So although
we have two dilaton equations, we have just {\it one}\,
dilaton operator $\ddg {1,1}$.

\section{Landau-Ginzburg representation}

\setcounter{equation}{0}
\setcounter{subsection}{0}

In the $c<1$ models the integrable structure
provides a quite effective way to compute the correlation
functions and the latter admit a topological Landau-Ginzburg
interpretation, \cite{eguchi1}. Now we are going to show that this is also true
in the $c=1$ case. We will first exhibit very general formulas to
calculate the correlation functions. In the full coupling space it is
very difficult to solve the
$W$--constraints exactly, so as to obtain explicit expressions for the
correlators. However in some subspaces, the calculation is drastically
simplified. In such cases it will be possible to explicitly see
how a Landau-Ginzburg interpretation shows up.

\subsection{Correlation functions on subspace ${\cal S}$}

In this section we will consider the subspace ${\cal S}$, i.e.
we require all the extra couplings to vanish (except $g_{1,1}=-1$).
The dispersionless coupling conditions (\ref{discoupling})
will reduce to
\a
M =\tl, \qquad
L=\hs(\tm). \label{discouplings}
\b
and
\a
\ddg {i,j} M(\zeta) + iL^{i-1}\tl^j=\ddg {i,j} \tl(\zeta), \qquad
\ddg {i,j} \hs(\tm(\zeta))) + jL^i\tl^{j-1}=\ddg {i,j} L(\zeta).
\label{discouplings'}
\b
Starting from eq.(\ref{wc}), taking suitable derivatives with respect
to flow parameters, then restricting to ${\cal S}$,
we obtain
\ai
<\chi_{n,m}> &=&\frac{1}{(n+1)(m+1)}\oint M^{m+1}(L)dL^{n+1}, \label{1pchis}\\
<\chi_{k,l}\chi_{n,m}> &=& \frac{1}{(n+1)(m+1)}
\ddg {k,l} \oint M^{m+1}(L)dL^{n+1}\0\\
&+&km(1-\delta_{l,0}) <\chi_{n+k-1, m+l-1}>, \label{2pchis}\\
<\chi_{r,s}\chi_{k,l}\chi_{n,m}> &=& \frac{1}{(n+1)(m+1)}
 \frac{\d^2}{\d g_{r,s} \d g_{k,l}} \oint M^{m+1}(L)dL^{n+1} \0\\
&+& km (1-\delta_{l0}) <\chi_{r,s}\chi_{n+k-1, m+l-1}>\0\\
&+& rm (1-\delta_{s0}) <\chi_{k,l}\chi_{n+r-1, m+s-1}> \label{3pchis}\\
&-&   rkm(m-1)(1-\delta_{l0})(1-\delta_{s0}) <\chi_{n+r+k-2, m+s+l-2}>. \0
\bj
The most general multi--point correlation functions in ${\cal S}$ have been
given in \cite{BX2}
\a
< \chi_{n,m} \prod_{\mu=1}^k \chi_{i_\mu,j_\mu} >
&=& \frac{1}{(n+1)(m+1)} \Bigl(\prod_{\mu=1}^k \ddg {i_\mu,j_\mu}\Bigl)
  \oint M^{m+1}(L)dL^{n+1} \0\\
&+&  \sum_{r=1}^k (-1)^{r+1} \frac{m!}{(m-r)!}
   \sum_{\rho\in S_k}\prod_{\mu=1}^r \Bigl(i_{\rho(\mu)}
   (1-\delta_{j_{\rho(\mu)}0})\Bigl)   \label{multipchis}\\
&\cdot &  < \chi_{n+i_{\rho(r+1)}+\ldots+i_{\rho(k)}-r,m+j_{\rho(r+1)}+\ldots
   j_{\rho(k)}-r} \prod_{s=1}^r \chi_{i_{\rho(s)},j_{\rho(s)}} >, \0
\b
where $\rho$ is an element of the symmetric group $S_k$. In all the above
formulas the contour integral is understood around $\infty$.

\subsection{The pure tachyonic sector in  ${\cal S}_+$}

The above equations are only formal, unless we are able to compute
the explicit expressions of $L$ and $\tl$. To know the latter we have to solve
the coupling conditions (\ref{discoupling}) for $L$ and $\tl$. This can be done
explicitly for the simplest ${\cal M}_{p,q}$ models, \cite{BCX}, but the
formulas are quite complicated. Since our purpose here is to unveil the
LG structure of the c=1 string theory, we will limit ourselves to
${\cal S}_+$ and ${\cal S}_-$.
Due to the  ${\bf Z}_2$ symmetry of Toda hierarchy, from now on,
we will only consider the case ${\cal S}_+$, the discussion for the
small space ${\cal S}_-$ being exactly specular.
In the parameter space ${\cal S}_+$,
the solution to  eq.(\ref{discouplings}) is very simple
\a
&& R = x, \qquad a_0 = t_{2,1}, \qquad a_l = 0, \qquad \forall l\geq2 \0\\
&& b_i = x^i \sum_{r\geq i+1} r\left(\begin{array}{c} r-1 \\ i
 \end{array}\right)
  t_{2,1}^{r-i-1}t_{1,r}, \qquad \forall i\geq0, \label{solution}
\b
or equivalently
\a
 L = \zeta + t_{2,1} \qquad
 \tl= \frac{x}{\zeta} + \sum_{r=1}^\infty rt_{1,r} L^{r-1}. \label{solultl}
\b
Plugging these expressions into eq.(\ref{multipchis}), we can get
all the correlation functions explicitly. We are now ready to introduce a LG
representation of $c=1$ string theory. It consists of picking a potential $W$
and representatives for the fields, and showing that they
satisfy the properties of a LG topological field theory in such a way that
we can identify it with ${\cal T}_1$. As will be apparent in a moment,
the potential we have to choose is $W=\tl$, which is non--polynomial in
$\zeta$.
The representatives of the fields  will be denoted $\phi_{r,s}$. They are
to be identified later on with $\chi_{r,s}$, but, for the sake of
clarity,  we prefer to keep the two symbols distinct.

Let us, for the time being,
restrict our attention to the pure tachyonic sector. We define
\a
 \phi_n\equiv\phi_{n,0}\equiv (L^n)'_+, \qquad
\phi_{-m}\equiv\phi_{0,m}\equiv -(\tl^m)_-', \qquad
\phi_0=\phi_{0,0}=\frac{1}{\zeta}, \qquad n,m\geq1. \label{phii,j}
\b
Then, using eqs.(\ref{ddtim'}--\ref{ddgijtm'}), we can simplify
the formula for three point function (\ref{3pchis}), and get
\a
<\phi_a \phi_b\phi_c>=-\oint_{\zeta=0}\,
\frac{\phi_a\phi_b\phi_c}{\tl'}. \label{3tachyons}
\b
where $a,b,c$ are integers.
The LHS represents the correlation functions of three tachyons.
The other multi--point tachyon correlation functions can be obtained
by simply taking derivatives with respect to additional couplings,
for example, the four-point function is
\a
<T\phi_a\phi_b\phi_c>=-\frac{\d}{\d t}
\oint_{\zeta=0}\frac{\phi_a\phi_b\phi_c}{\tl'}.\label{4tachyons}
\b
where $t$ represents either $t_{1,n}(n\geq1)$ or $t_{2,m}(m\geq1)$,
or $x$, accordingly to whether $T$ is $T_n, T_{-m}$ or $Q$.

With the above identifications, the residue formula (\ref{3tachyons})
is the same as in the more well--known $c<1$ Landau--Ginzburg models, except
for one detail.
In the standard  Landau--Ginzburg theory, the integral
contour surrounds all the zeroes of the superpotential, while in the present
case
it surrounds the origin. In ${\cal S}_0$ the two contour integrals coincide
since the only poles of the integrand can be at zero and at $\infty$, but
in general this equivalence has to be verified.

The residue formula (\ref{3tachyons}) suggests that $\phi_a$ are the
representatives of primaries of a topological LG theory. Let us find
further confirmations of this suggestion. To this end
let us consider the restricted
integrable hierarchy. Eqs.(\ref{ddtim'}, \ref{ddgijm'}) imply that
\a
\frac{\d L(\zeta)}{\d t_i}=0, \qquad
\frac{\d L(\zeta)}{\d x}=0, \qquad
\frac{\d L(\zeta)}{\d \tt_1}=1, \0
\b
and
\a
\frac{\d \tl(\zeta)}{\d t_i}=\phi_{i}, \qquad
\frac{\d \tl(\zeta)}{\d x}=\phi_0, \qquad
\frac{\d \tl(\zeta)}{\d \tt_1}=\phi_{-1}+\tl',
\b
where we have used the fact that $\tl=M$ in ${\cal S}_+$.
These equations imply that the only non-vanishing contacts between
the primary fields and others are
\ai
\frac{\d\phi_{-j} (\zeta)}{\d t_{1,i}}&=& \Bigl[\frac{\phi_{i}
  \phi_{-j}}{\tl'}\Bigl]'_-, \\
\frac{\d\phi_{-j} (\zeta)}{\d x}&=& \Bigl[\frac{\phi_0
  \phi_{-j}}{\tl'}\Bigl]'_-, \\
\frac{\d\phi_{i} (\zeta)}{\d t_{2,1}}&=& \phi_{i}', \\
\frac{\d\phi_{-j} (\zeta)}{\d t_{2,1}}&=& \phi_{-j}' +
\Bigl[\frac{\phi_{-1}
  \phi_{-j}}{\tl'}\Bigl]'_-.
\bj
In particular we see that $\phi_{1}$ has vanishing contacts
with all the other primary fields, and lowers the level
of the gravitational descendants by one, i.e.
\a
\frac{\d\phi_{-j} (\zeta)}{\d t_{1,1}}=j \phi_{-j+1}.
\b
This confirms our identifications of the puncture operator
and the primary fields.

\subsection{The discrete states}

Now let us turn our attention to the discrete states. We still work
with ${\cal S}_+$. The representatives of the discrete states
are defined by the Laurent series
\a
 \phi_{i,j}\equiv iL^{i-1}\tl^j -(L^i\tl^j)_-'
  =(L^i\tl^j)_+'-jL^i\tl^{j-1}\tl', \qquad
\phi_0=\phi_{0,0}=\frac{1}{\zeta}. \label{phiij}
\b
For pure tachyons this coincides with the defintion in the
previous subsection. One may wonder why we choose such a peculiar
combination (which is not a total derivative w.r.t.
$\zeta$, unlike the usual situation in $c<1$ case). This is
uniquely determined by the requirement that the three point function
have a residue formula expression. Before giving a proof,
let us derive the restricted flow equations of $L, \tl$ and $M$ in
${\cal S}_+$. Using eqs.(\ref{distoda1}, \ref{distoda2}) and
(\ref{ddtim'}--\ref{ddgijtm'}), we get
\ai
\frac{\d L(\zeta)}{\d g_{i,j}}& =& j(L^i\tl^{j-1})_{\leq0},
    \label{dldgijs+}\\
\frac{\d M(\zeta)}{\d g_{i,j}} &=& \delta_{j0}(L^i)'- j(L^i\tl^j)'_-
 + j(L^i\tl^{j-1})_{\leq0}M'(\zeta),   \label{dmdgijs+}\\
\frac{\d \tl(\zeta)}{\d g_{i,j}} &=& iL^{i-1}\tl^j - j(L^i\tl^j)'_-
 + j(L^i\tl^{j-1})_{\leq0}M'(\zeta). \label{dtldgijs+}
\bj
They lead to
\a
\frac{\d \tl(\zeta)}{\d g_{i,j}}=\phi_{i,j}  +
\frac{\d L(\zeta)}{\d g_{i,j}}M'(\zeta). \label{eqm}
\b
The flow equations of the fields $\phi_{i,j}$ in ${\cal S}_+$ constitute a
part of the so--called {\it contact algebra},
\ai
\frac{\d\phi_{k,l} (\zeta)}{\d t_{1,i}} &=& \frac{ikl}{i+k-1}\phi_{i+k-1,l-1}
  +\frac{i-1}{i+k-1}\Bigl[\frac{\phi_{i,0}\phi_{k,l}}{\tl'}\Bigl]'_-,
  \0\\
\frac{\d\phi_{k,l} (\zeta)}{\d x} &=& \Bigl[\frac{\phi_0\phi_{k,l}}
  {\tl'}\Bigl]'_--\phi_0 \Bigl(\frac{\phi_{k,l}}{\tl'}\Bigl)'_-
  +l\phi_0\phi_{k,l-1}, \label{contactalgti}\\
\frac{\d\phi_{k,l} (\zeta)}{\d t_{2,1}} &=& \phi_{i,j}'
 +\Bigl[\frac{\phi_{0,1}\phi_{k,l}}
  {\tl'}\Bigl]'_-+\phi_{0,1} \Bigl(\phi_{i,j-1}-j
  \frac{\phi_{k,l}}{\tl'}\Bigl)'_-.\0
 \bj
In particular, we have
\a
&& \frac{\d\phi_{k,l} (\zeta)}{\d t_1} = l\phi_{k,l-1}, \qquad
  \{k\geq1,l\geq0\}\oplus \{k=0,l\geq2\}; \label{punctures+}\\
&& \frac{\d\phi_{0,1} (\zeta)}{\d t_1} = 0, \qquad
\frac{\d\phi_0 (\zeta)}{\d t_1} =0. \0
\b
This once again confirms that $\ddt {1,1}$ is indeed a puncture operator.
Furthermore, the flow equations of $L, \tl, M$, and $\phi_{i,j}$
enable us to derive the simplified formulas for multi--point
 correlation functions.
Then the first few multi--point functions are
\a
<\phi_{n,m}> &=& \frac{1}{(n+1)(m+1)}\oint\, \tl^{m+1} dL^{n+1}, \\
<\phi_{k,l}\phi_{n,m}> &=& \oint\, (L^k\tl^l)_- d(L^n\tl^m)\0\\
  &+& \frac{km(1-\delta_{l0})}{(n+k)(m+l)}\oint\, \tl^{m+l} dL^{n+k},\\
<\phi_{i,j}\phi_{k,l}\phi_{n,m}>
&=&-\oint\, \frac{\phi_{i,j}\phi_{k,l}\phi_{n,m}}{\tl'}. \label{3ptlg}
\b
In the derivation of the residue formula for the three point function,
we have used eq.(\ref{3pchis}), and the flow equations. However,
if the correlators contains at least one primary, we
can have a simpler derivation. Let us start from eq.(\ref{h2pf}),
take one more derivative w.r.t. the coupling parameter, and make
use of the equations of motion, we have
\a
&&~~<\phi_{i,j}\phi_{k,l}\phi_{n,0}>
=\ddg {k,l} \oint (L^i\tl^j)_- dL^n
= \oint \Bigl[ \phi_{n,0} \frac{\d (L^i\tl^j)_-}{\d g_{k,l}}
-(L^i\tl^j)_-' \frac{\d (L^n)}{\d g_{k,l}}\Bigl]d\zeta \0\\
&&= \oint d\zeta\phi_{n,0}\Bigl[
  + jL^i\tl^{j-1}\phi_{k,l} + l(L^k\tl^{l-1})_{\leq0}\Bigl(
   iL^{i-1}\tl^j+jL^i\tl^{j-1}M'
  -(L^i\tl^j)_-'\Bigl)\Bigl]\0\\
&&= \oint d\zeta\phi_{n,0}\Bigl[jL^i\tl^{j-1}\phi_{k,l}
  + l(L^k\tl^{l-1})_{\leq0}(L^i\tl^j)_+'\Bigl]\0\\
&&= -\oint \frac{ \phi_{n,0}\phi_{i,j}\phi_{k,l}}{\tl'}
  + \oint d\zeta\phi_{n,0}(L^i\tl^j)_+'  \Bigl[
  l(L^k\tl^{l-1})_{\leq0}+\frac{\phi_{k,l}}{\tl'}\Bigl]\0\\
&&= -\oint \frac{ \phi_{n,0}\phi_{i,j}\phi_{k,l}}{\tl'}.\0
\b
In the third step we have used the equality $M'=\tl'$, in the last step
we have used the fact that
\a
\oint d\zeta\frac{ f_+(\zeta)}{\tl'}=0, \qquad \forall f.\0
\b
Since we have used the equality $M'=\tl'$, in general the four--point
functions are not obtainable by simply taking derivative w.r.t.
the additional coupling parameter. But this is true if the fourth
parameter is the coupling to tachyon or the cosmological constant, i.e.
\a
<T\phi_{i,j}\phi_{k,l}\phi_{n,m}>
=-\frac{\d}{\d t}
 \oint\, \frac{\phi_{i,j}\phi_{k,l}\phi_{n,m}}{\tl'}. \label{4gtachyons}
\b
where $t$ represents either $t_{1,i}(i\geq1)$ or $t_{2,j}(j\geq1)$,
or  $x$, according to whether $T$ is either $T_i, T_{-j}$ or $Q$.
This is further evidence that all tachyons are primary fields.

The rationale behind the construction of this subsection is as follows.
Since the primary fields span the most general Laurent series of $\zeta$,
the gravitational descendants $\phi_{i,j}$ are particular combinations
of the primary fields. Therefore any correlation function involving
gravitational descendants can be expressed in terms of the
correlation functions among only the primary fields.

\subsection{Unperturbed LG }

So far we have been working on ${\cal S}_+$. The formulas we have obtained are
very suggestive of a LG framework, however they may look a bit involved
especially at a first reading. For this reason, in this subsection
we consider an even simpler situation, the coupling space ${\cal S}_0$,
where the LG interpretation is particularly clear. In this case,
\a
W(\zeta)= \frac{x}{\zeta}, \qquad
\phi_{i,j}(\zeta) = M(i,j) x^j \zeta^{i-j-1}, \label{phiijs0}
\b
which shows that fields are highly degenerated, for
\a
\phi_{i+n,i}(\zeta) = (n+i) x^i\phi_{n,0}, \qquad
\phi_{j,j+m}(\zeta) = -(m+j) x^j\phi_{0,m}.
\b
Therefore, all the correlation functions (containing discrete states)
can be expressed in terms of the correlation functions among
tachyons
\ai
<\phi_{i+n,i}\phi_\alpha\phi_\beta>
&=&\frac{(n+i)}{ni}<\phi_{i,0}\phi_{0,i}><\phi_{n,0}\phi_\alpha\phi_\beta>, \\
<\phi_{j,m+j}\phi_\alpha\phi_\beta>
&=&-\frac{(m+j)}{mj}<\phi_{j,0}\phi_{0,j}><\phi_{0,m}\phi_\alpha\phi_\beta>.
\bj
One can easily compute
the multi--point correlation functions by means of (\ref{1pchis}--\ref{3pchis})
or (\ref{3ptlg}). The results are those of section 2.  From the above formulas
one can easily prove once again the puncture equations and recursion relations
within the LG formalism, and extract the algebra ${\cal R}_1$ which was
introduced in section 2.1.

\centerline{----------------}

In this section we have shown that the extended $2d$ dispersionless Toda
hierarchy subject to proper constraints and in the pure cosmological sector
${\cal S}_0$, admits a topological Landau-Ginzburg
formulation exactly similar the $c<1$ models. In a larger
coupling space however some of the typical equations, such as the puncture
equations and recursion relations, do not have in general exactly the same
form as the $c<1$ models. In such a case the correct form of these relations
is embodied in the flow equations of the dispersionless Toda hierarchy
and the relevant coupling conditions.

\section{Conclusions}

We think we can safely conclude that ${\cal T}_0$, ${\cal T}_1$
and ${\cal T}_{-1}$ are topological field theories both before and after
perturbation by all the tachyonic operators. They have an infinite set of
primaries; this seems to be an intrinsic characteristic. One may in fact ask
oneself whether we can truncate one of the above theories so as to obtain
a TFT with a finite number of primaries (truncation means fixing a
subset of primaries and keeping only
the correlators among these primaries). The answer is however negative.
One can extract from each of the above three theories infinitely many subsets
containing a finite number of fields such that the metrics are invertible, but
one easily realizes that associativity requires an infinite number of fields.
Therefore, although one can envisage many topological subtheories of
${\cal T}_0$, ${\cal T}_1$ and ${\cal T}_{-1}$, they must all contain an
infinite number of primaries.

There is a way to obtain submodels of the above TFT's, but it is far more
sophisticated than a simple truncation and can be best understood in the
framework of two--matrix model: one constrains the theory to live in a
particular submanifolds of the coupling space. For example, if, after switching
on the couplings $t_{1,1},t_{1,2},t_{1,3},t_{2,1},t_{2,2}$, one examines the
theory along
the direction $t_1\sim x$ -- the values of the remaining parameters is
actually irrelevant -- then one finds, \cite{BCX}, that the correlators
of $T_{2r+1}$ are the correlators of pure topological gravity and obey
the flow equations of the KdV hierarchy. More complicated submanifolds of
the coupling space generate the other KdV models and hierarchies. We quoted
the KdV case because it may help us understand the nature of the two puncture
operators $T_0$ and $T_1$. Since $T_1$ is conjugate to
$t_1$ while $T_0$ is conjugate to $x$, in the submanifold $t_1\sim x$ the
two operators collapse to the same object, which becomes the puncture
operator considered in \cite{WDVV}.

In this complicated but significant manner the TFT studied in this paper,
with its double nature, contains well--known TFT's coupled to topological
gravity.
\vskip1cm
{\bf Acknowledgements}. One of us (C.S.X.) would like to thank T.Eguchi for
helpful discussions and JSPS for financial support.

\end{document}